\documentclass[conference]{IEEEtran}

\usepackage[T1]{fontenc}

\usepackage{amsmath,amssymb,amsfonts}
\usepackage{algorithmic}
\usepackage{graphicx}
\usepackage{textcomp}
\usepackage[dvipsnames]{xcolor}

\usepackage[linesnumbered,ruled,vlined]{algorithm2e}
\usepackage{array}
\usepackage{url}
\usepackage[utf8]{inputenc} 
\usepackage{bm} 
\usepackage{caption}
\usepackage{subcaption}
\usepackage{booktabs} 
\usepackage{longtable} 
\usepackage{arydshln} 
\usepackage{xifthen} 
\usepackage{cite}
\usepackage{multirow}

\usepackage{arydshln}
\usepackage{soul}

\usepackage{stfloats}
\usepackage{float}

\usepackage{hyperref}

\graphicspath{{figure/}}

\newcommand{\thR}[0]{\alpha_{\textnormal{H}}}
\newcommand{\thE}[0]{\alpha_{\textnormal{L}}}

\newcommand{\abs}[1]{\left|#1\right|}

\newcommand{\ifconditon}[0]{\mathsf{B}} 

\newcommand{\wH}[0]{\mathsf{W_H}} 
\newcommand{\dH}[0]{\mathsf{d_H}} 

\newcommand{\df}[0]{N_{\textnormal{fh}}} 
\newcommand{\imax}[0]{I_{\max}}
\newcommand{\Lmax}[0]{L_{\max}}

\newcommand{\pch}[0]{\mathsf{p_{ch}}}
\newcommand{\pehm}[0]{\mathsf{p_a}} 
\newcommand{\pchminv}[0]{\mathsf{p_b}} 

\newcommand{\currentItr}[0]{^{\mathsf{r},(I)}}
\newcommand{\lastItr}[0]{^{\mathsf{c},(I-1)}}
\newcommand{\iniItr}[0]{^{\mathsf{c},(0)}}

\newcommand{\cmrb}[0]{_{0,[0]}}
\newcommand{\clrb}[0]{_{0,[1]}}
\newcommand{\elrb}[0]{_{1,[1]}}
\newcommand{\emrb}[0]{_{1,[0]}}

\begin{document}
	\title{Iterative Bounded Distance Decoding With 2-Bit Reliability Marker For Product-Like Codes}
	\author{%
		\IEEEauthorblockN{Xiaowei Wu\IEEEauthorrefmark{1}, Min Qiu\IEEEauthorrefmark{2}, Yueying Zhan\IEEEauthorrefmark{1}, and Lei Yang\IEEEauthorrefmark{1}}		
		\IEEEauthorblockA{
			\IEEEauthorrefmark{1}Technology and Engineering Center for Space Utilization, Chinese Academy of Sciences, Beijing, China \\ 
			\IEEEauthorrefmark{2}Global College, Shanghai Jiao Tong University, Shanghai, China. \\
			Email: \{wuxiaowei,zhanyueying,yang.lei\}@csu.ac.cn,  min\_qiu@sjtu.edu.cn}
		\thanks{This research was supported in part by the National Natural Science Foundation of China under Grant 62301533, 62671374, 62371439, and U24A20214; in part by the National Key R$\textnormal{\&}$D Program of China under Grant 2025YFF0522600; in part by the Prospective Foundation of CSU-CAS under Grant T4034711QY and T4035511QY. The work of Min Qiu was supported in part by the Shanghai Jiao Tong University ExploreX Funding under Grant SD6040004/153.}
	}
	\maketitle
	\begin{abstract}
		We propose a novel soft-aided hard-decision decoding scheme for product codes based on two-bit quantized reliability markers. In particular, one bit is used for a two-level tolerance miscorrection check to validate the bounded distance decoding (BDD) result, while the other identifies bits to be randomly flipped upon BDD failure and miscorrection. List decoding is incorporated to further facilitate the error performance. Density evolution is performed for the design of the quantization thresholds. Compared to iterative BDD, the proposed scheme requires only moderate increases in memory and computational overhead. Simulation results verify that our scheme outperforms several unquantized soft-aided decoders, while maintaining small memory costs and low BDD attempts.
	\end{abstract}
	
	\section{Introduction}   
	Product codes (PCs) \cite{productcode} and their variations, such as staircase codes \cite{Smith-staircase} and zipper codes \cite{Sukmadji-zipper}, are powerful forward error correction codes that are widely adopted for high-speed fiber-optic communications (e.g., 400ZR/800ZR \cite{ofec}) and considered potentially applicable for resource-constrained free-space optical communications \cite{viasat-fec}. While iterative bounded distance decoding (iBDD) offers the low complexity necessary for ultra-high throughput, its performance significantly lags behind turbo product decoding (TPD) \cite{tpd}. However, TPD relies on list decoding with a large number of test patterns as well as high-precision soft-message passing, leading to excessive internal data flow and implementation complexity.
	
	To bridge the gap between iBDD and TPD, soft-aided hard-decision decoding (SA-HDD) \cite{Liga-sabm,Sheikh-ibddsr,Sheikh-ibddcr,Sheikh-beepc,Rapp-eed,Miao-drsd,Miao-drsd2,Rapp-drsd3,Lei-isabm,Li-ibddrf} has emerged as a promising hybrid approach. SA-HDDs retain bounded distance decoding (BDD) as the core decoding component and leverage soft channel information to aid the BDD result, achieving a favorable trade-off between performance and complexity. One typical approach of SA-HDDs is to treat PCs as generalized low-density parity-check (GLDPC) codes. These schemes leverage channel log-likelihood ratios (LLRs) to derive soft reliability for BDD outputs, as seen in iBDD with scaled reliability (iBDD-SR) \cite{Sheikh-ibddsr}, iBDD with combined reliability (iBDD-CR) \cite{Sheikh-ibddcr}, and iBDD with random flipping (iBDD-RF) \cite{Li-ibddrf}. Another common approach is to assign reliability markers/scores based on magnitudes of soft channel output. These markers are used for miscorrection checks (MCC) and/or bit-flipping across iterations, as seen in soft-aided bit marking (SABM) decoding \cite{Liga-sabm,Lei-isabm}, error-and-erasure decoding (EaED) \cite{Rapp-eed}, and dynamic reliability score decoding (DRSD) \cite{Miao-drsd,Miao-drsd2,Rapp-drsd3}. Most notably, DRSD \cite{Miao-drsd} and its extensions \cite{Miao-drsd2,Rapp-drsd3} have demonstrated error performance approaching TPD. Despite this progress, SA-HDDs face several challenges for practical implementations. Particularly, various schemes use unquantized or highly quantized soft information, leading to high memory costs. 
	
	To tackle this challenge, we propose a low complexity SA-HDD scheme using 2-bit reliability markers. Our scheme adopts the first-MCC-then-flipping workflow, specifically designed to exploit 2-bit reliability markers with minimal numerical computation. Moreover, a two-level miscorrection tolerance is introduced to reduce the probability of error propagation due to false identification of high-reliability bits during the iterative decoding process. We derive the density evolution equations to optimize the quantization thresholds. Simulation results demonstrate that the proposed scheme achieves significant coding gains over iBDD and several benchmark SA-HDDs with unquantized soft information, while maintaining small memory costs and low BDD attempts.
	
	\emph{Notations}: Scalars, vectors, and matrices are written in light, small bold, and capital bold letters, respectively, e.g., $x$, $\bm{x}$, and $\mathbf{X}$. $\abs{x}$ denote the absolute value of $x$. The $i$-th element in a vector $\bm{x}$ is denoted by $x_i$. $\wH(\bm{x})$ denotes the Hamming weight of binary vector $\bm{x}$. $\dH(\bm{x},\bm{y})$ denotes the Hamming distance between binary vectors $\bm{x}$ and $\bm{y}$. $\oplus$ and $\otimes$ denote XOR and AND operators, respectively. $\ifconditon(\mathsf{cond})$ is a hard-decision operator that outputs either 1 or 0 to indicate whether the given condition ``$\mathsf{cond}$'' is satisfied or not. When ``$\mathsf{cond}$'' is a vector of $n$ conditions, the output is a length-$n$ binary vector. $\mathsf{rand}\{0,1\}^n$ returns an $n$-bit random binary sequence. 
	
	\section{Product Codes and iBDD}	
	Let $\mathcal{C}(n, k, t)$ be a BCH code with codeword length $n$, information length $k$, and designed error correction capability $t$. We consider a product code where $\mathcal{C}$ is employed as both the row and column component code, resulting in a code rate of $R=k^2/n^2$. When a PC codeword $\mathbf{X}=[x_{i,j}]$ of size $n \times n$ is transmitted, the corresponding soft and hard channel output matrices are denoted by $\mathbf{R}=[r_{i,j}]$ and $\mathbf{Y}=[y_{i,j}]$, respectively.
	We consider an additive white Gaussian noise (AWGN) channel with binary inputs. For a transmitted bit $x_{i,j}\in \{0,1\}$, the soft channel output is $r_{i,j}=(-1)^{1-x_{i,j}} + z_{i,j}$, where $z_{i,j}\sim \mathcal{N}(0,\sigma^2)$ is the real-valued Gaussian noise with zero mean and variance $\sigma^2 = 1/(2 R E_b/N_0)$. The hard-decision channel output is obtained as $y_{i,j}=\ifconditon(r_{i,j}>0)$. 
	
	The component BCH codes are decoded using bounded distance decoding (BDD). Let $\bm{x}$ be a BCH codeword being transmitted and $\bm{y}$ be the received sequence. The BDD decoder declares a decoding success if there exists a codeword $\hat{\bm{x}}\in\mathcal{C}$ that is no greater than $t$ bits away from $\bm{y}$, i.e., $\dH(\hat{\bm{x}},\bm{y}) \leq  t$. In this case, the BDD result could be either a valid correction ($\hat{\bm{x}}=\bm{x}$) or a miscorrection ($\hat{\bm{x}}\neq\bm{x}$). If no such codeword exists, the decoder declares a decoding failure and outputs the original received sequence $\hat{\bm{x}}=\bm{y}\notin\mathcal{C}$.
	
	The received hard-decision PC block is processed using iBDD with intrinsic message passing (IMP), which alternately processes the rows and columns of the block with BDD. The iteration terminates when either all rows and columns have zero syndrome or a maximum number of iterations is reached. 
	
	\section{Proposed Decoder}	
	The proposed two-bit reliability-marker-aided iBDD (iBDD-RM2) scheme utilizes two binary markers. When a soft PC block $\mathbf{R}$ is received, a 3-bit quantization is performed, producing the hard-decision PC block $\mathbf{Y}$ and two binary reliability marker matrices, denoted by $\mathbf{H}$ and $\mathbf{L}$. Following each BDD attempt, the decoder either accepts the BDD output or executes a bit-flipping operation with the help of reliability markers. Furthermore, the proposed scheme can incorporate list decoding via randomized BDD retries
	to achieve additional coding gains.
	
	\subsection{Reliability Marker Generation}
	For each received bit $r_{i,j}\in \mathbf{R}$ and its corresponding hard-decision $y_{i,j}$, a reliability score $S(y_{i,j})$ is assigned using
	\begin{align}
		S(y_{i,j})=\left\lbrace \begin{aligned}
			& 2 \text{ (high)} && \text{if } \abs{r_{i,j}} \geq \thR \\
			& 1 \text{ (medium)} && \text{if } \thE \leq \abs{r_{i,j}} < \thR \\
			& 0 \text{ (low)} && \text{if } 0\leq\abs{r_{i,j}} < \thE
		\end{aligned}
		\right. ,
	\end{align}
	where $(\thE,\thR)$ are the quantization thresholds. The reliability score $S(y_{i,j})$ is expressed in the form of a pair of reliability markers $S(y_{i,j})=(h(y_{i,j}),\ell(y_{i,j}))$, given by
	\begin{align}
		\left\lbrace \begin{aligned}
			& h(y_{i,j}) \triangleq \ifconditon (\abs{r_{i,j}} \geq \thR) \\
			& \ell(y_{i,j}) \triangleq \ifconditon (\abs{r_{i,j}} < \thE) 
		\end{aligned}
		\right. , 
	\end{align}
	so that $\mathbf{H}=[h(y_{i,j})]$ and $\mathbf{L}=[\ell(y_{i,j})]$. 	
	The high reliability bit (HRB) marker $\mathbf{H}$ is used for MCC and not to be updated throughout the iterative decoding process. The low reliability bit (LRB) marker $\mathbf{L}$ is used for bit flipping and is updated depending on the validity of BDD outputs. 
	
	Given the quantization thresholds $(\thE,\thR)$, the capacity of the 3-bit quantized channel $W:\{x\}\rightarrow\{(y, h(y), \ell(y))\}$ is 
	\begin{align}
		& C(\sigma,\thE,\thR) =  \nonumber \\
		& \quad \sum_{S=0}^{2} P_{e,S} \log_2 \frac{2P_{e,S}}{P_{e,S}+P_{c,S}} + 
		P_{c,S} \log_2 \frac{2P_{c,S}}{P_{e,S}+P_{c,S}} ,
	\end{align}
	where $P_{e,S}$/$P_{c,S}$ is the probability of an erroneous/correct hard-decision with reliability score $S$, expressed as
	\begin{subequations}\label{eq:qfunc1}\begin{align}
			& \left\lbrace \begin{aligned}
				& P_{e,2} \!=\! Q\left(\frac{1+\thR}{\sigma}\right)  \\
				& P_{e,1} \!=\! \pehm\!-\!P_{e,2} \\
				& P_{e,0} \!=\! \pch\!-\!\pehm 
			\end{aligned} \right. ,\\
			& \left\lbrace \begin{aligned} 
				& P_{c,2} \!=\! 1\!-\!Q\left(\frac{1-\thR}{\sigma}\right) \\
				& P_{c,1} \!=\! P_{c,2}\!-\!\pchminv \\
				& P_{c,0} \!=\! \pchminv\!-\!\pch 
			\end{aligned} \right. ,\\
			& \left\lbrace \begin{aligned} 
				& \pch \!=\! Q\left(\frac{1}{\sigma}\right) \\
				& \pehm \!=\! Q\left(\frac{1+\thE}{\sigma}\right) \\
				& \pchminv \!=\! Q\left(\frac{1-\thE}{\sigma}\right)
			\end{aligned} \right. ,
	\end{align}\end{subequations}  
	and $Q(\cdot)$ refers to the Q-function. With properly designed thresholds, this 3-bit quantized channel can have a significant capacity gain over the hard-decision binary input AWGN channel and the EaED channel in \cite[Sec. II-B-1]{Rapp-eed}.

	\subsection{Two-bit Reliability Marker Aided iBDD}
	The proposed iBDD-RM2 employs IMP \cite{Rapp-eed} for the iterative decoding of a PC block $\mathbf{Y}$. The soft-aided decoding for a row or column component $\bm{y}$ involves three steps: 1) BDD decoding, 2) applying MCC aided by HRB markers, and 3) updating LRB markers and flipping the LRBs. Optionally, list decoding is considered. The list can be processed either serially or in parallel without affecting the error performance. Similar to most existing SA-HDD schemes, two iBDD iterations are conducted after finishing the soft-aided iterations.
	
	\subsubsection{iBDD-RM2 with serial list decoding}
	The detailed procedures for the proposed scheme with serial list decoding are given in Alg. 1. In step 1, the BDD inputs a row/column $\bm{y}=[y_1, \dots ,y_n]$ of the received PC block and generates the decoding result $\hat{\bm{x}}=[\hat{x}_{1},\dots,\hat{x}_{n}]$, together with an error flag $E$ and an error vector $\bm{e}$, expressed as
	\begin{align}
		\left\lbrace \begin{aligned}
			&E=\ifconditon(\hat{\bm{x}}\notin\mathcal{C}) , \\
			&\bm{e}=\bm{y}\oplus\hat{\bm{x}} \label{eq:bdd_e2} .
		\end{aligned} \right.
	\end{align}
	
	\begin{figure}[t]
		\centering
		\vspace{1pt}
		\includegraphics[width=\columnwidth]{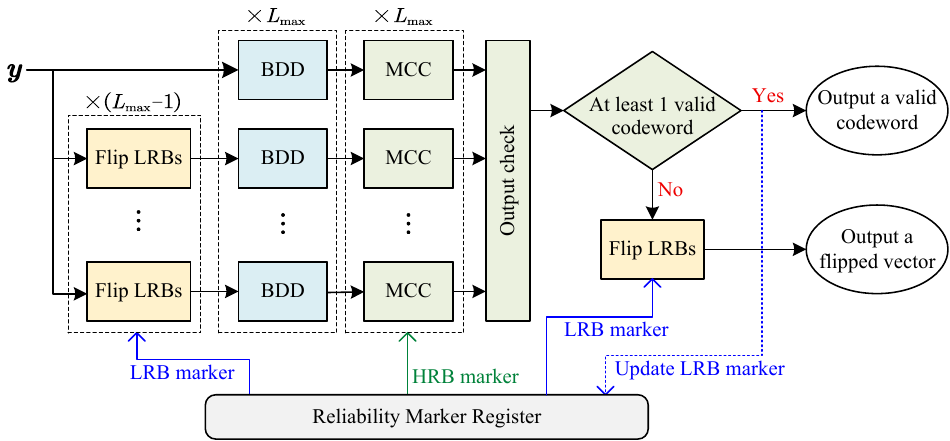}
		\caption{2-bit reliability marker aided BDD with parallel list. }
		\label{fig:com_dec}		
	\end{figure}
	
	In step 2, the HRB markers $\bm{h}=[h(y_1),\dots,h(y_n)]$ are used to validate the BDD output, producing $(V,\df{})$ using MCC based on $(E,\bm{e},\bm{h},T)$:
	\begin{align}
		&\df{}=\wH(\bm{e} \otimes \bm{h})= \sum_{i=1}^{n} e_i \cdot h(y_i) , \label{eq:mcc1} \\
		&V=\left\lbrace \begin{aligned}
			& 0 										&& \text{if } E=1 \\
			&\ifconditon(\df{} \leq T) && \text{if } E=0 \\
		\end{aligned} \right.  , \label{eq:mcc2} 
	\end{align}
	where $V$ is the validity flag and $\df{}$ is the number of conflicts between HRBs and the BDD error vector. When BDD fails, the output is immediately considered invalid $(V=0,\df{}=0)$ without carrying on the above computations. When BDD  succeeds, the output is identified as a miscorrection if the number of flipped HRBs exceeds the MCC tolerance $T$, which is initialized to $0$ for the first half of the soft iterations and updated to $1$ for the remaining iterations. 
	
	In step 3, LRB markers $\bm{\ell}=[\ell(y_1),\dots,\ell(y_n)]$ are updated according to MCC results. Specifically, $\bm{\ell}$ is unflagged if the BDD gives a valid correction with no conflict $(V=1,\df{}=0)$. For a valid BDD output, the decoder sets $\bm{w}=\hat{\bm{x}}$. Otherwise, the LRBs in $\bm{y}$ are flipped with a probability of 1/2 using a bit-wise selector $\bm{w}=\mathsf{flip}(\bm{y},\bm{\ell})$ defined as
	\begin{align}
		\begin{aligned}
			{w}_{i} = \left\{\begin{aligned}
				&\mathsf{rand}\{0,1\} &&  \textrm{if } \ell(y_i)=1\\
				&{y}_{i} && \textrm{otherwise}
			\end{aligned} \right. 
		\end{aligned} , \label{eq:fliplrb} 
	\end{align}
	for $i=1,\dots,n$. Last, $\bm{w}$ is output as the decoding result. 
	
	With a decoding list size of $\Lmax \geq 2$, steps 1-3 are repeated up to $\Lmax$ times. Starting from the second iteration, the BDD input is replaced by the sequence $\bm{w}$ produced in the previous iteration. This inner iteration calls early termination once a valid BDD output $\bm{w}$ is given. 	
	
	\begin{algorithm}[h!]
		\caption{2-bit reliability marker aided BDD\\ with serial list decoding of list size $\Lmax$} \label{alg:sabdd_serial}
		\small
		\DontPrintSemicolon
		\KwIn{$\bm{y}$, $\bm{h}$, $\bm{\ell}$, $\Lmax$}
		\KwOut{$\bm{w},\bm{\ell}$}
		\For{$L=1:\Lmax$}{		
			\If {$L=1$} {$(\hat{\bm{x}},E,\bm{e})\leftarrow \mathsf{BDD}(\bm{y})$ \tcp{see \eqref{eq:bdd_e2}}} 
			\Else {$(\hat{\bm{x}},E,\bm{e})\leftarrow \mathsf{BDD}(\bm{w})$} 
			$T\leftarrow \ifconditon(\text{current\_iteration}>\imax/2) $\;
			$(V,\df{})\leftarrow\mathsf{MCC}(E,\bm{e},\bm{h},T)$ \tcp{see \eqref{eq:mcc1}-\eqref{eq:mcc2}}
			\If{$V=0$}{
				$\bm{w}\leftarrow\mathsf{flip}(\bm{y},\bm{\ell})$ \tcp{see \eqref{eq:fliplrb}}
			}
			\Else{		
				\If{$\df{}=0$}{$\bm{\ell} \leftarrow \mathbf{0}$}  
				$\bm{w}\leftarrow\hat{\bm{x}}$ \;
				\textbf{break}
			}
		}
		\Return{$\bm{w}, \bm{\ell}$}
	\end{algorithm}
	
	\subsubsection{iBDD-RM2 with parallel list decoding}
	To mitigate the decoding latency associated with large list sizes, the proposed scheme can be implemented using a parallel architecture, as depicted in Fig. \ref{fig:com_dec}. For a list size $\Lmax \geq 2$, the first BDD decoder processes the original received sequence $\bm{y}$, while the remaining $\Lmax-1$ BDD decoders simultaneously process distinct variants of $\bm{y}$, obtained by the random flipping of LRBs. If one or more BDD decoders produce valid results, the first valid result is selected as the final soft-aided output. When all BDD decoders fail to return a valid output, the LRBs are randomly flipped again to produce the soft-aided output. 
	
	A dynamic list size can be employed to further reduce the average number of decoding attempts at high SNR. Given a target maximum list size $\Lmax$, the list size is set to 1 in the first iteration and the immediate iteration after MCC tolerance update. In subsequent iterations, the list size is doubled at each step until $\Lmax$ is reached.

	\subsection{Optimization of Quantization Thresholds $(\thE,\thR)$}
	The quantization thresholds $(\thE,\thR)$ are selected respectively according to the component BCH codes. Since they are the only configurable parameters, their optimal values can be determined using a simple two-step numerical search. 
	
	First, we use the assumption that all BDD miscorrections can be detected. Under this condition, the decoding result for a received sequence $\bm{y}$ becomes 
	\begin{align}
		\hat{\bm{x}}=\left\lbrace \begin{aligned}
			& \bm{x}&& \text{if } \dH(\bm{x},\bm{y}) \leq t \\
			& \mathsf{flip}(\bm{y},\bm{\ell}) &&  \text{if } \dH(\bm{x},\bm{y}) > t \\
		\end{aligned} \right. .		
	\end{align}	
	Numerical search is executed with this assumption to find the LRB quantization threshold $\thE$ that minimizes the SNR required for reaching a BER of $10^{-4}$ with ${\imax}=10$ iterations and $\Lmax=1$. Next, with the obtained $\thE$, a numerical search is performed using our proposed scheme with the same decoder configuration to optimize the HRB quantization threshold $\thR$. 
	
	We conducted this numerical search for $n=127,255,511$ and $t=2,3,4$. As shown in Fig. \ref{fig:the_threshold}, the optimized $\thE$ decreases as either $n$ or $t$ increases. Meanwhile, the optimized $\thR$ rises with $t$ and is relatively insensitive to the variation of $n$. 
	
	\subsection{Density Evolution Analysis and $\thE$ Optimization} 
	
	We also investigate the optimization of the LRB threshold $\thE$ using density evolution (DE) over extrinsic message passing (EMP), following the analysis framework in \cite{7954697,Sheikh-ibddsr}. 
	Consider the row-decoding phase in the $I$-th iteration, where the message passing can be expressed by \eqref{eq:emp_msg}. For $i'\in\{1,\dots,n\}\setminus i$, the $i'$-th variable node (VN) passes the decoding message $\psi_{i',j}\lastItr$ to the $j$-th constraint node (CN). Meanwhile, the $i$-th VN passes the hard-decision channel output $\hat{r}_i$ to the $j$-th CN. The decision message given by the $j$-th CN and forwarded to the $j^\prime$-th CN via the $i$-th VN is denoted by $\psi_{i,j^\prime}\currentItr$. 
	
	\begin{figure}[t]
		\centering
		\vspace{1pt}
		\includegraphics[width=\columnwidth]{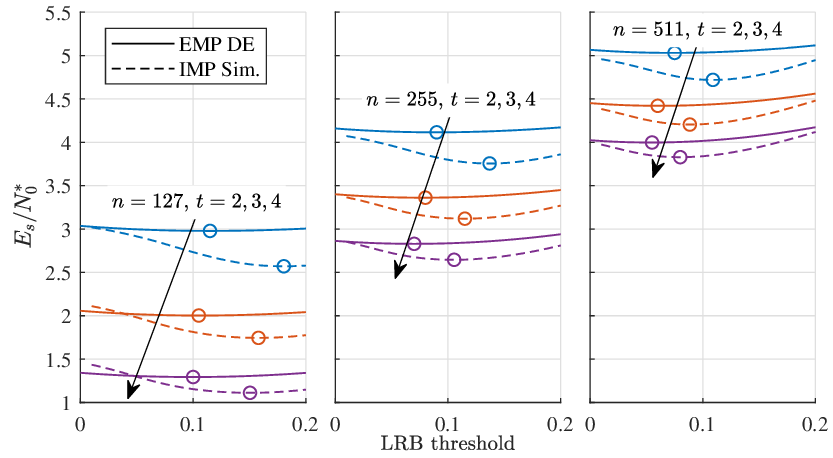}
		\caption{SNR threshold $E_s/N_0^\ast(\thE)$, at which the BER of $10^{-4}$ is reached with the given $\thE$. ``$\circ$'' marks optimized $\thE$. ${\imax}=5$ for EMP DE, and ${\imax}=10$ for IMP simulation. }
		\label{fig:the_threshold}		
	\end{figure}
	
	\begin{figure*}[ht]
		\begin{align}
			\text{VNs}(1,\dots,n) \xrightarrow[]{\left[\psi_{1,j}\lastItr,\dots, \psi_{i-1,j}\lastItr,\hat{r}_i,\psi_{i+1,j}\lastItr,\dots, \psi_{n,j}\lastItr\right]}  \text{CN}(j) \xrightarrow[]{\psi_{i,j^\prime}\currentItr} \text{VN}(i) \xrightarrow[]{\psi_{i,j^\prime}\currentItr} \text{CN}(j^\prime) .  \label{eq:emp_msg}
		\end{align}		 
		\hrule 
	\end{figure*}
	
	To perform DE, we assume transmitting an all-zero codeword (due to the symmetric channel) and an ideal BDD. The VN output $\psi_{i,j^\prime}\currentItr $ is decided by the two-step approach: 
	\begin{align}
		&\hspace{-3pt}\mu_{i,j^\prime}\currentItr =\left\lbrace \begin{aligned}
			& 0 && \text{if BDD succeed}   \\
			& \emptyset && \text{if BDD failed} 
		\end{aligned} \right. ,   \\
		&\hspace{-3pt}
		\psi_{i,j^\prime}\currentItr \!=\! \left\lbrace \begin{aligned}
			& 0 && \!\!\text{if } \mu_{i,j^\prime}\currentItr \!=\! 0 \\
			& \hat{r}_i && \!\!\text{if } \mu_{i,j^\prime}\currentItr \!=\! \emptyset \text{ and } \ell(\hat{r}_i)\lastItr = 0 \\
			& \text{\small $\mathsf{rand}\{0,1\}$} && \!\!\text{if } {\mu_{i,j^\prime}\currentItr } \!=\! \emptyset \text{ and } \ell(\hat{r}_i)\lastItr \!=\! 1 \\
		\end{aligned} \right. .	\hspace{-3pt}
	\end{align}	
	The LRB marker $\ell(\hat{r}_i)\lastItr$ is initialized as $\ell(\hat{r}_i)\iniItr = \ifconditon( |\hat{r}_i| < \thE)$ in the first iteration, and updated after BDD attempt using 
	\begin{align} 
		\ell(\hat{r}_i)\currentItr \!=\! \left\lbrace \begin{aligned}
			& 0 \qquad \text{if }  \mu_{i,j^\prime}\currentItr \!=\! 0 \text{ and }   \ell(\hat{r}_i)\lastItr \!=\! 1 \\
			& \ell(\hat{r}_i)\lastItr \quad \text{otherwise}
		\end{aligned} \right. .
	\end{align}
	
	Let $\mathtt{x}\currentItr$ be the error probability of $\psi_{i,j^\prime}\currentItr$ and $\mathtt{x}\lastItr$ be the error probability of $\psi_{i',j}\lastItr$ for $i'\in\{1,\dots,n\}\setminus i$. In the first iteration, the error probability $\mathtt{x}\iniItr$ is initialized to the error probability of the hard-decision channel output, given by $\pch \triangleq Q\left(1/\sigma\right)$. The derivation of $\mathtt{x}\currentItr$ is as follows. 
	
	Let $f_{\emptyset|\hat{r}_i}(\mathtt{x})$ be the probability of BDD failure when $\hat{r}_i$ is given, while knowing that the error probability of the other $n-1$ input decoding messages is $\mathtt{x}$.  For a component code with blocklength $n$ and error-correcting capability $t$, these terms are calculated as follows:
	\begin{subequations} \label{eq:de_1}\begin{align}
			f_{\emptyset|0}(\mathtt{x}) 
			& = \sum_{j=t+1}^{n-1} \binom{n-1}{j} \mathtt{x}^j  (1-\mathtt{x})^{n-1-j} , \\
			f_{\emptyset|1}(\mathtt{x}) 
			& = \sum_{j=t}^{n-1}  \binom{n-1}{j} \mathtt{x}^j  (1-\mathtt{x})^{n-1-j} .
	\end{align}\end{subequations}	
	The joint probability $P_{a,[b]}\lastItr\triangleq\Pr\{\hat{r}_i=a,\ell(\hat{r}_i)\lastItr=b\}$ is initialized in the first iteration by following \eqref{eq:qfunc1}
	\begin{align}
		\left\lbrace \begin{aligned}
			& P\cmrb\iniItr \!=\! 1-\pchminv  \\
			& P\clrb\iniItr \!=\! p_{c,0}  \\
			& P\elrb\iniItr \!=\! p_{e,0}  \\
			& P\emrb\iniItr \!=\! \pehm  
		\end{aligned} \right. .
	\end{align}
	After LRB marker updating,  ${P_{a,[b]}\currentItr\!\triangleq\!\Pr\{\hat{r}_i\!=\!a,\ell(\hat{r}_i)\currentItr\!=\!b\}}$ is obtained by 
	\begin{subequations}\begin{align}
			\left\lbrace \begin{aligned}
				&P\cmrb\currentItr=P\cmrb\lastItr+P_{\text{uf},0} \\ &P\clrb\currentItr=P\clrb\lastItr-P_{\text{uf},0} 
			\end{aligned} \right. ,	\\
			\left\lbrace \begin{aligned}
				&P\elrb\currentItr=P\elrb\lastItr-P_{\text{uf},1}\\
				&P\emrb\currentItr=P\emrb\lastItr+P_{\text{uf},1}
			\end{aligned} \right. ,
	\end{align}\end{subequations}
	where $P_{\text{uf},\hat{r}_i}$ is the probability that the LRB marker $\ell(\hat{r}_i)\currentItr$ is un-flagged, expressed as
	\begin{subequations} \label{eq:de_2} \begin{align}
			&P_{\text{uf},0}=P\clrb\lastItr \cdot (1-f_{\emptyset|0}(\mathtt{x}\lastItr)) , \\
			&P_{\text{uf},1}=P\elrb\lastItr \cdot (1-f_{\emptyset|1}(\mathtt{x}\lastItr)) . 
	\end{align}\end{subequations}
	
	Employing Bayes’ rule, $\mathtt{x}\currentItr$ can be written as 	
	\begin{align}
		&\begin{aligned}
			\mathtt{x}\currentItr =&P\emrb\lastItr \cdot f_{\emptyset|1}(\mathtt{x}\lastItr)+ \\
			& 0.5 \cdot \sum_{b\in\{0,1\}} \left( P_{b,[1]}\lastItr \cdot f_{\emptyset|b}(\mathtt{x}\lastItr)  \right) , 
		\end{aligned} 
	\end{align}
	where each term is given in \eqref{eq:de_1}-\eqref{eq:de_2}. The DE equations for column decoding are omitted as they can be derived in the same manner as in the row decoding phase. 
	
	Fig. \ref{fig:the_threshold} compares the optimization results of $\thE$ obtained via numerical search (IMP simulation) and EMP DE analysis given above. The $\thE$ derived using EMP DE is slightly lower than the one from IMP simulation, yet they demonstrate a consistent decreasing trend for increased $n$ or $t$.  
	
	\subsection{Update of MCC tolerance} \label{sec:mcc}
	
	\begin{figure}[t]
		\centering				
		\subcaptionbox{Weight-3 error \label{fig:error_pattern_w5}}{\includegraphics[width=4cm]{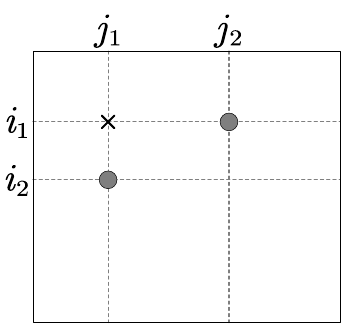}}
		\hspace{5pt}
		\subcaptionbox{Weight-6 error \label{fig:error_pattern_w6}}{\includegraphics[width=4cm]{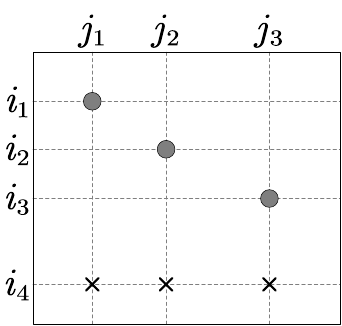}}		
		\caption{Low-weight error patterns of PCs with $t=2$ due to error-HRBs. ``$\bullet$'' denotes error-HRBs and ``$\times$'' denotes erroneous LRBs/MRBs.} 
		\label{fig:error_pattern}
	\end{figure}	
	
	Product decoders that employ miscorrection checks based on anchor bits  \cite{Anchor} or similar high-reliability indicators \cite{Miao-drsd,Liga-sabm} can be inherently vulnerable to performance degradation when these bits are erroneous. In the proposed iBDD-RM2, the adoption of a two-level MCC tolerance $T$ is critical, particularly when the component BCH code has ${t=2}$. To be specific, the miscorrection rate of iBDD decreases over decoding iterations. This can be verified using the DE analysis for iBDD-SR \cite[Sec. IV-A]{Sheikh-ibddsr} by setting the scaling factor to $\infty$ for all iterations. Hence, to mitigate error propagation due to miscorrection and prevent false alarms of error bits identified as HRBs (error-HRBs), a tight MCC tolerance $T=0$ is initialized to mitigate error propagation in the early stages, and subsequently loosened to $T=1$ after  $\imax/2$ iterations.
	
	\begin{figure}[t]
		\centering
		\subcaptionbox{$P_{\text{ue}}$ with ${\Lmax\!=\!1}$ and at $E_b/N_0$ of 4.6 dB}{\includegraphics[width=7.5cm]{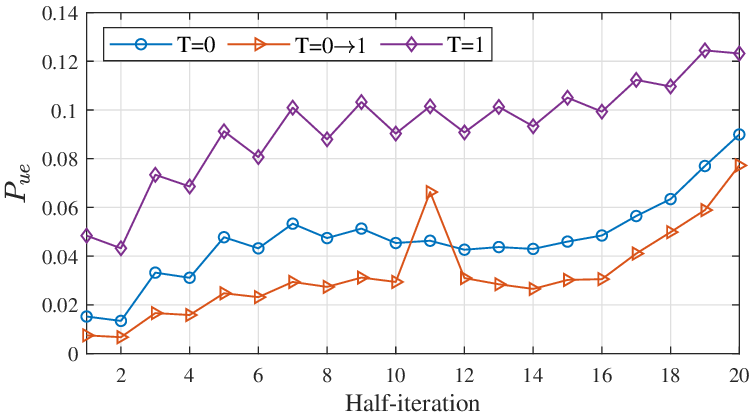}}
		\subcaptionbox{$P_{\text{fa}}$ with ${\Lmax\!=\!1}$ and at $E_b/N_0$ of 4.6 dB}{\includegraphics[width=7.5cm]{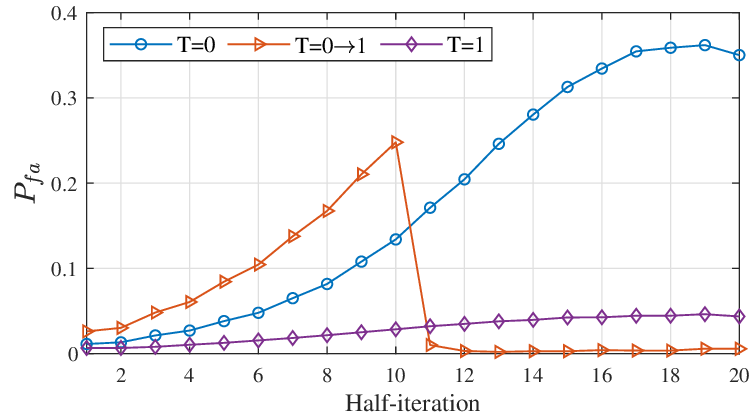}}
		\subcaptionbox{BER performance}{\includegraphics[width=7.5cm]{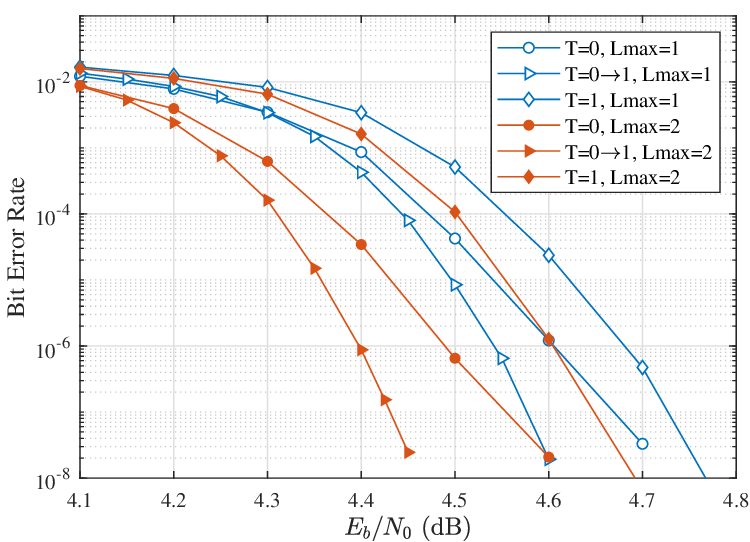}}
		\caption{Performance metrics for different MCC tolerances with $\mathcal{C}(255,239,2)$ and ${{\imax}\!=\!10}$. The HRB thresholds $\thR$ are optimized as 0.6, 0.5, and 0.3 for ${T\!=\!0}$, ${T\!=\!0 \!\to\! 1}$, and ${T\!=\!1}$, respectively. }  \label{fig:misscorrRate}
	\end{figure}
	
	To depict this, consider a PC with $t=2$ BCH component codes and a fixed MCC tolerance of $T=0$ is used for decoding. Fig. \ref{fig:error_pattern} plots two examples of typical low-weight error patterns involving error-HRBs. Although both error patterns are correctable if a standard iBDD is used, the error-HRBs lead to MCC false alarms when a fixed MCC tolerance of $T=0$ is used. In consequence, valid corrections of BDD are rejected. As an identified miscorrection is always followed by random flipping of LRBs, such cases could induce more errors in later iterations. While the final iBDD stages can mitigate low-density errors, multiple error-HRB-induced patterns may jointly trigger severe error propagation that exceeds the error correction capability of the final iBDD stages. 
	
	Due to this observation, the MCC tolerance $T$ is loosened to 1 for the second half of the soft iterations to avoid decoding failure caused by error-HRBs. This allows BDD corrections that only flip 1 HRB to be accepted as valid results. Clearly, once the tolerance loosens, the error patterns similar to the examples shown in Fig. \ref{fig:error_pattern} can be corrected within one iteration. 
	
	Figs. \ref{fig:misscorrRate}a-b compare the evolution the undetected error rate $P_{\text{ue}}$ and false alarm rate $P_{\text{fa}}$ over decoding iteration for a PC with $\mathcal{C}(255,239,2)$. To be specific, $P_{\text{ue}}$ and $P_{\text{fa}}$ represent the ratio of accepted miscorrections and rejected valid corrections to total MCC attempts, respectively. 
	When fixing ${T\!=\!0}$, both $P_{\text{ue}}$ and $P_{\text{fa}}$ increase with iterations. Particularly, $P_{\text{fa}}$ exceeds $20\%$ after the 6-th iteration. On the contrary, fixing ${T\!=\!1}$ leads to a persistently high $P_{\text{ue}}$ in all iterations, while having low $P_{\text{fa}}$. 
	For ${T\!=\!0 \!\to\! 1}$, the tolerance update diminishes $P_{\text{fa}}$ to nearly zero within one iteration. Although this triggers a marginal increment of $P_{\text{ue}}$, it remains lower than that of fixed-tolerance in subsequent iterations.
	The error performance for different MCC tolerances is shown in Fig. \ref{fig:misscorrRate}c. The two-level tolerance significantly outperforms fixed tolerance. This performance gain becomes more pronounced as the list size increases.
	
	\section{Numerical Results and Discussions} 
	
	Monte-Carlo simulations are conducted to evaluate the performance of iBDD-RM2. BCH codes used to construct PCs and the quantization thresholds, which are fixed regardless of the channel SNR, are summarized in Table \ref{tb:quan_optimized}. 
	
	\begin{table}[h]
		\centering
		\caption{Quantization thresholds for PCs} \label{tb:quan_optimized}
		\begin{tabular}{ccccc}
			\toprule
			$\mathcal{C}$ & $\mathcal{C}_1$(255,239,2) & $\mathcal{C}_2$(255,238,2) & $\mathcal{C}_3$(255,231,3)\\ 
			\midrule
			$\thE$ & 0.13 & 0.13 & 0.12 \\  
			$\thR$ & 0.5 & 0.5 & 0.6 \\  
			\bottomrule
		\end{tabular}
	\end{table}
	
	\subsection{Error performance}
	
	Fig. \ref{fig:ber_i10} shows the BER performance of iBDD-RM2, together with benchmark schemes iBDD-SR \cite{Sheikh-ibddsr}, iBDD-CR \cite{Sheikh-ibddcr}, iBDD-RF \cite{Li-ibddrf}, and DRSD \cite{Miao-drsd}, as reported in their respective original papers. iBDD-RM2 and iBDD-SR/CR/RF use ${\imax\!=\!10}$ soft-aided iterations and $2$ additional iBDD iterations. iBDD-SR/CR/RF use unquantized soft information. DRSD uses a 5-bit reliability score and ${\imax\!=\!10}$/$20$ soft-aided iterations with list size ${\Lmax\!=\!2}$. It can be seen that the proposed iBDD-RM2 outperforms iBDD-SR/CR/RF with unquantized soft information and approaches the performance of DRSD. 
	
	\begin{figure}[t]
		\centering	
		\subcaptionbox{}{\includegraphics[width=7.5cm]{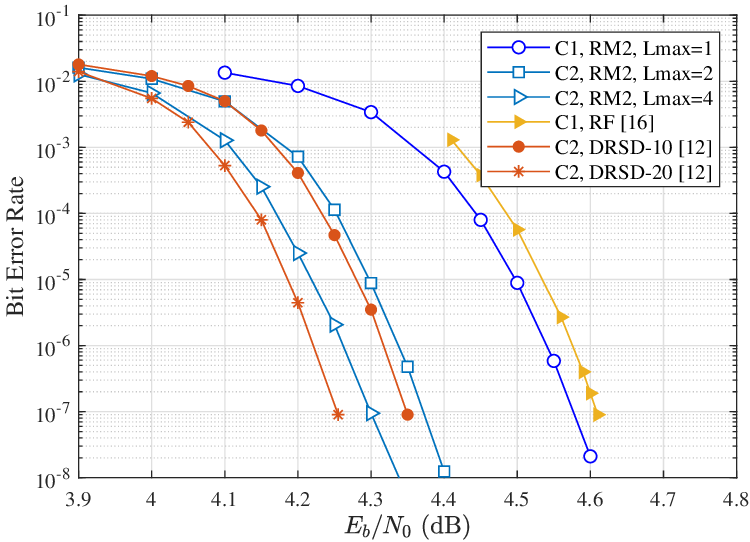}}
		\subcaptionbox{}{\includegraphics[width=7.5cm]{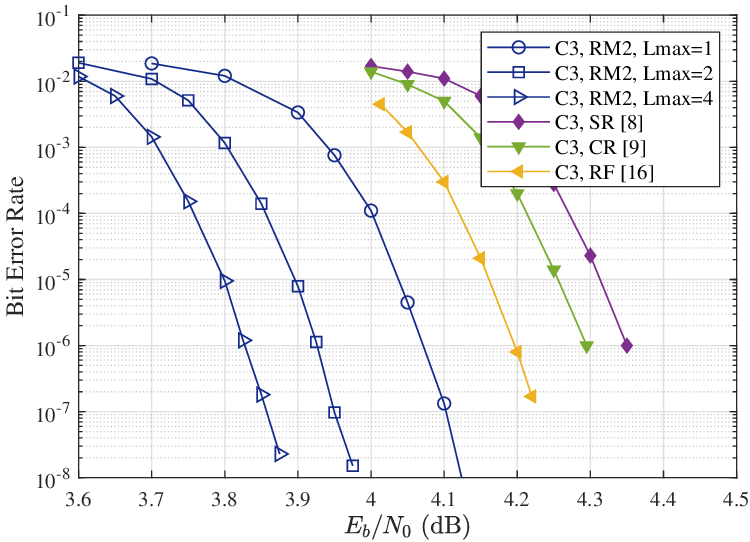}}
		\caption{BER performance of iBDD-RM2 for PCs with components $\mathcal{C}_1$(255,239,2),  $\mathcal{C}_2$(255,238,2), and $\mathcal{C}_3$(255,231,3) at ${\imax}=10$ iterations. As a benchmark, iBDD with 20 maximum iterations reaches a BER of $10^{-6}$ at 5.36, 5.06, and 4.64 for $\mathcal{C}_1$ - $\mathcal{C}_3$, respectively. } \label{fig:ber_i10}
	\end{figure}
	
	\subsection{Memory Costs and Decoding Complexity}
	We evaluate the additional memory costs of the proposed iBDD-RM2, iBDD-SR/CR/RF, and DRSD, when compared to iBDD. As summarized in Table \ref{tb:memory}, the memory costs arising from soft information are measured using dynamic reliability metrics (DRM), static reliability metrics (SRM), lookup tables (LUT), and quantization resolution (QR). First, iBDD-SR, iBDD-CR, and iBDD-RF need to store static channel LLRs and the scaling factor lookup table. In addition, iBDD-RF needs to store a bit-flipping probability lookup table. DRSD uses $q$-bit reliability scores (feasible with at least 3-bit quantization) and 1-bit erasure markers for decoding, and both are updated dynamically. In contrast, iBDD-RM2 requires only two binary reliability markers $(\mathbf{H},\mathbf{L})$, where $\mathbf{H}$ is static and $\mathbf{L}$ is dynamic. Hence, iBDD-RM2 shows a significant memory benefit compared to most benchmark decoders. 
	
	\begin{table}[t]
		\caption{Additional memory costs of SA-HDDs.} \label{tb:memory}
		\setlength{\tabcolsep}{5pt}
		\centering
		\begin{tabular}{lccccc}
			\toprule
			& iBDD-RM2 & iBDD-SR & iBDD-CR & iBDD-RF & DRSD \\
			\midrule
			{DRM} & $n^2$ & - & - & - & $(q+1)n^2$\\ 
			{SRM} & $n^2$ & $qn^2$ & $qn^2$ & $qn^2$ &  \\ 
			LUT & - & $2q\imax $  & $6q\imax $   & $4q\imax $  & - \\
			\midrule
			{QR} & $q=2$ & $q \geq 1$  &$q \geq 3$  & $q \geq 2$ & $q \geq 3$  \\
			\bottomrule 	
			\multicolumn{6}{p{0.9\linewidth}}{\vspace{-4pt}$\ast$ $q$: quantization bit length of soft-information.}\\
			\multicolumn{6}{p{0.9\linewidth}}{$\ast$ DRM: dynamic reliability metrics that can be updated.}\\
			\multicolumn{6}{p{0.9\linewidth}}{$\ast$ SRM: static reliability metrics that are fixed in the decoding process.}\\
			\multicolumn{6}{p{0.9\linewidth}}{$\ast$ LUT: parameters (such as scaling factors or bit-flipping probabilities) stored in a lookup table wherein different entries are accessed during each iteration.}\\
			\multicolumn{6}{p{0.9\linewidth}}{$\ast$ QR: quantization resolution demands for soft-information.}\\
			\bottomrule
		\end{tabular}
	\end{table}%
	\begin{figure}[t]
		\centering
		\subcaptionbox{Normalized BDD attempts.	\label{fig:complexity_bdd} }{\includegraphics[width=7.5cm]{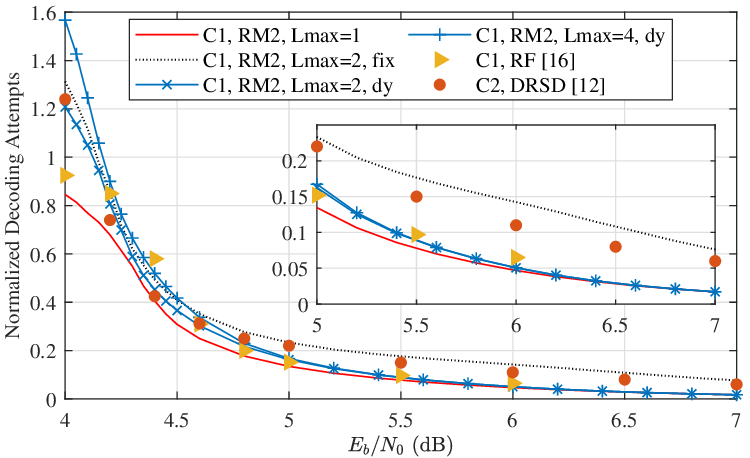}}
		\subcaptionbox{Normalized MCC attempts  \label{fig:complexity_mcc} }{\includegraphics[width=7.5cm]{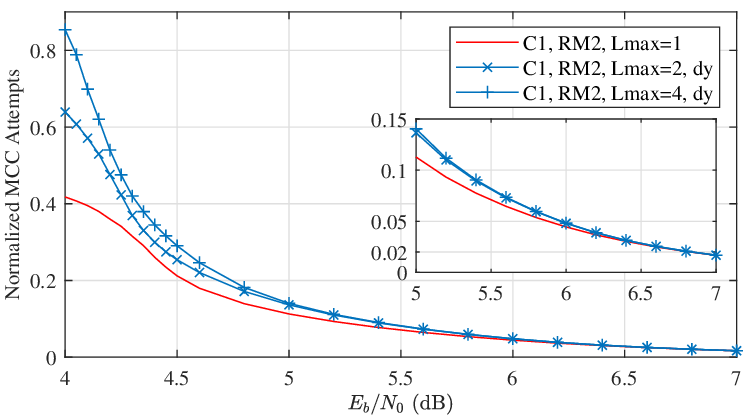}}	
		\caption{Complexity metrics for iBDD-RM2. $\imax = 10$ for all results. iBDD-RM2 with $\Lmax=2$ is implemented with parallel list, where ``fix'' denotes fixed list size and ``dy'' denotes dynamic list size. } 
		\label{fig:complexity}
	\end{figure}
	
	The primary computational overhead of SA-HDD schemes comes from the increased number of BDD decoding attempts. To evaluate this, Fig. \ref{fig:complexity}a plots the average number of BDD decoding attempts (including the final iBDD iterations) normalized by $2n\imax$, for iBDD-RM2, iBDD-RF, and DRSD. Compared to iBDD-RF, iBDD-RM2 achieves better BER with fewer decoding attempts. Compared to iBDD-RM2 with a fixed list size,  iBDD-RM2 using a dynamic list size reduces the decoding attempts and nearly halves the decoding attempts at high SNRs. 
	
	The MCC operator also contributes to the computation overhead. In Fig. \ref{fig:complexity}b, we plot the number of MCC attempts, normalized by $2n\imax$. Approximately $50\%$-$60\%$ of the decoding attempts invoke MCC attempts at low SNRs, whereas only a moderate number of MCCs are performed at high SNRs. 
	
	The overhead associated with updating LRB markers is relatively small. Unlike DRSD, where reliability scores are updated in every iteration, the proposed scheme updates LRB markers at most once. Specifically, the only updating operation is to unflag the LRB markers from 1 to 0 when the BDD generates a highly confident output with $(V=1,\df{}=0)$.

	\section{Conclusions} 
	This paper introduced iBDD-RM2, a novel 2-bit reliability-marker-aided BDD decoding scheme for product codes. By employing a two-level MCC tolerance and flexible list decoding, the scheme enhances coding gains with low computation overhead. Simulation verified that iBDD-RM2 significantly outperforms baseline iBDD and several unquantized SA-HDD benchmarks. Its superior performance-complexity trade-off makes it a promising candidate for high-speed optical communications on resource-constrained platforms.

	\bibliographystyle{ieeetr}
	\bibliography{references_ibdd}
	
	\newpage
	\appendices
	\section{Supplementary for Sec. IV}
	
	\subsubsection{$\mathcal{C}_1(255,239,2)$ versus $\mathcal{C}_2(255,238,2)$}
Fig. \ref{fig:c1c2} plots the BER curves of PCs with component codes  $\mathcal{C}_1$ and 	$\mathcal{C}_2$, respectively. To be specific, $\mathcal{C}_2$ is the even-weight subcode \cite{Miao-drsd} of $\mathcal{C}_1$. One can see that the PC with $\mathcal{C}_2$ outperforms the PC with $\mathcal{C}_1$. In addition, the PC with $\mathcal{C}_2$ benefits more from larger decoding lists. 

\subsubsection{Random flipping of LRBs}
To implement the list decoding with $\Lmax > 1$, candidate variants for the BDD input sequence $\bm{y}$ are generated based on the Hamming weight of the low-reliability bit (LRB) marker vector, denoted as $\wH(\boldsymbol{\ell})$:
\begin{itemize}
	\item ${\wH(\boldsymbol{\ell}) = 1}$: The single identified LRB is flipped, generating exactly one candidate variant.
	\item ${\wH(\boldsymbol{\ell}) = 2}$: The flipping patterns for the two LRBs are selected from the non-zero set $\{(0,1), (1,0), (1,1)\}$, producing up to three unique candidate variants, so that duplicate patterns are inherently avoided.
	\item ${\wH(\bm{\ell}) \geq 3}$: The flipping patterns are generated randomly and independently. In this case, duplicate candidates may occasionally occur.
\end{itemize}
Duplicate candidates only lead to a slight loss in error performance. For instance, for $\text{PC-}\mathcal{C}_2$ with $\imax = 10$ and $\Lmax = 4$, allowing duplicate candidates incurs a minor performance loss of less than $0.03$ dB at a BER of $10^{-6}$.

\subsubsection{BER gain over iBDD}
Fig.~\ref{fig:BER_n127n255n511} shows the performance gain of iBDD-RM2 over iBDD across component code lengths $n \in \{127, 255, 511\}$ and error-correcting capabilities $t \in \{2, 3, 4\}$. The corresponding optimized quantization thresholds for these component codes are summarized in Table~\ref{tb:th_configs}. Across all evaluated code parameters, iBDD-RM2 offers substantial gain over iBDD, confirming the effectiveness of our design.

\subsubsection{Per-Half-Iteration Performance}
Fig.~\ref{fig:decNum_perItr} depicts the number of BDD attempts per half-iteration (normalized by $n$). In general, BDD attempts drop over iterations as the residual errors are resolved. During the final iBDD iterations, only a few BDD attempts are invoked.

\begin{table}[H]
	\centering
	\caption{Quantization thresholds $(\thE,\thR)$ of iBDD-RM2 for PCs with various component codes. }
	\label{tb:th_configs}
	\begin{tabular}{cccc}
		\toprule
		& $n=127$ 	& $n=255$ 	  & $n=511$ \\
		\midrule
		$t=2$ & (0.18, 0.5) & (0.13, 0.5) & (0.11, 0.5) \\
		$t=3$ & (0.16, 0.6) & (0.12, 0.6) & (0.09, 0.6) \\
		$t=4$ & (0.15, 0.8) & (0.10, 0.8) & (0.08, 0.8) \\
		\bottomrule
	\end{tabular}
\end{table}
\vfill
\pagebreak

	\begin{figure}[H]
		\centering
		\includegraphics[width=7.5cm]{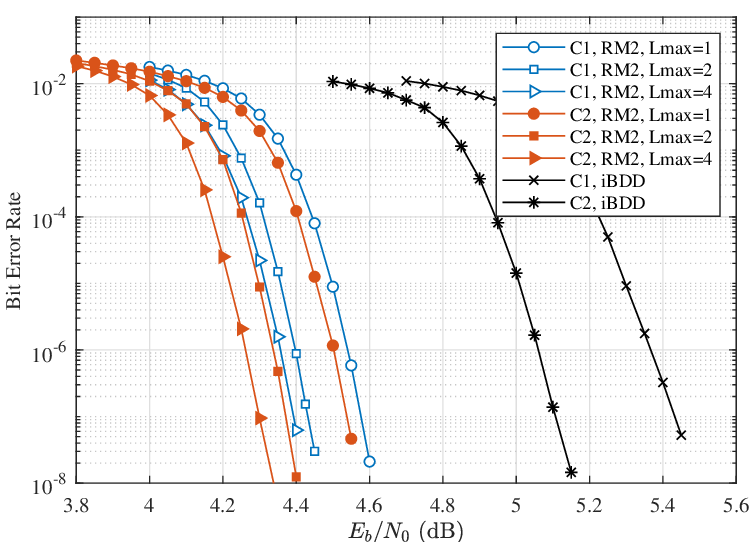}
		\caption{BER performance for PCs with components $\mathcal{C}_1$(255,239,2) and $\mathcal{C}_2$(255,238,2). iBDD-RM2 uses $\imax=10$ and $\Lmax=1,2,4$.} 
		\label{fig:c1c2}
		\vspace{5pt}
		\includegraphics[width=8.7cm]{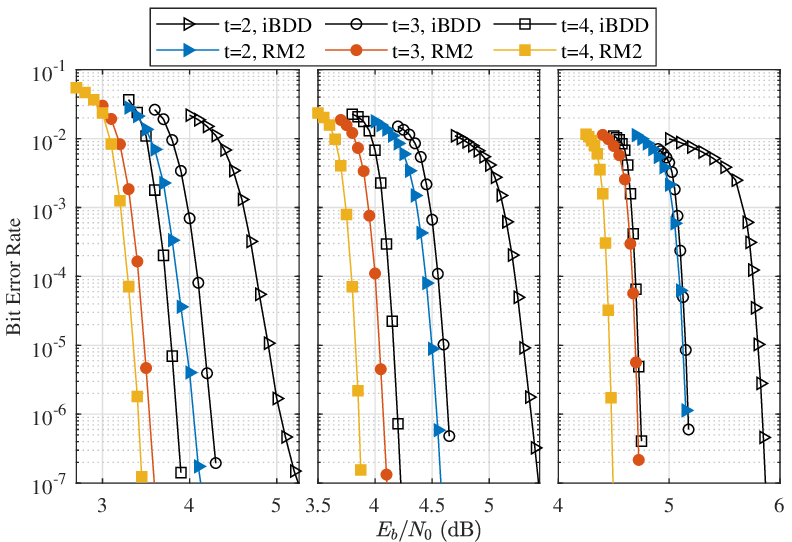}
		\caption{BER performance for PCs with component codes of $n=127,255,511$ and $t=2,3,4$.  iBDD-RM2 uses $\imax=10$ and $\Lmax=1$.} 
		\label{fig:BER_n127n255n511}
		\vspace{5pt}
		\includegraphics[width=7.5cm]{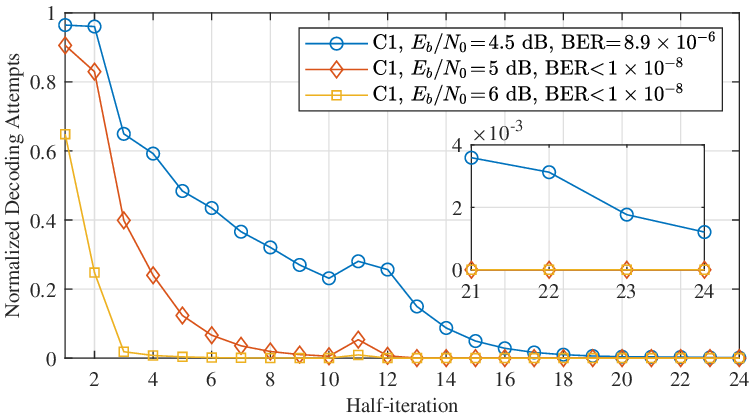}
		\caption{Normalized BDD attempts over half-iterations. } 
		\label{fig:decNum_perItr}
		\vfill
	\end{figure}
	\vfill
	
\end{document}